\documentclass{ws-p8-50x6-00}
\usepackage{epsfig,citesort,amsmath,amssymb,comment,a4p}
\usepackage{color}

\def\etal{{\em et al.\/}}

\def\tmw4j{$\mw4j$}

\def\be{\begin{equation}}
\def\ee{\end{equation}}
\def\bea{\begin{eqnarray}}
\def\eea{\end{eqnarray}}
\def\la{\label}
\def\bc{\begin{center}}
\def\ec{\end{center}}
\def\al{\langle}
\def\ar{\rangle}
\def\leq{\leqslant}

\def\ea{{\sl et al.}}
\def\eg{{\sl e.g.}}

\def\z{Z$^0$}
\def\pT{p_T}
\def\phi{\Phi}

\def\3d{y$$\times$$\phi$$\times$$\pT}

\def\OP{OPAL }

\def\col{Collaboration}

\def\PYTHIA{{\sc Pythia}}

\def\aver#1{\langle#1\rangle}

\newcommand{\ZF}[3]{Z. Phys. {\bf C{#1}} ({#2}) {#3}}
\newcommand{\ZP}[3]{Z. Phys. {\bf C{#1}} ({#2}) {#3}}

\newcommand{\NP}[3]{Nucl. Phys. {\bf {#1}} ({#2}) {#3}}
\renewcommand{\ZF}[3]{Z. Phys. {\bf C{#1}} ({#2}) {#3}}
\renewcommand{\ZP}[3]{Z. Phys. {\bf C{#1}} ({#2}) {#3}}

\renewcommand{\NP}[3]{Nucl. Phys. {\bf {#1}} ({#2}) {#3}}

\begin{document}

\title{ Genuine Correlations 
%of Like-Sign Particles \\ 
in   Hadronic {\z} Decays}

\author{E.A. De Wolf\\[1ex] for the OPAL Collaboration}

\address{CERN, European Organisation for Nuclear Research,
CH-1211 Geneva 23, Switzerland\\
 Physics Department, University of Antwerpen, B-2610 Antwerpen, Belgium  \\ 
E-mail: Eddi.DeWolf@ua.ac.be}

\maketitle

\abstracts{\noindent Correlations among hadrons with the same
electric charge produced in \z\ decays are studied using the high statistics  data
%a sample of over  4$\times 10^6$ multihadronic events
collected from 1991 through 1995 with the \OP detector at LEP.
Normalized factorial cumulants up to fourth order are used to measure genuine particle correlations 
as a function of the size of phase space  domains
 in rapidity, azimuthal angle and transverse momentum.
%   
%
%Both  all-charge  and like-sign particle combinations show
%strong positive genuine correlations.
%They are stronger in rapidity than in azimuthal angle.
%One-dimensional  cumulants initially 
%increase  rapidly  with decreasing size of the phase space cells
%but saturate quickly.
%The cumulants in two- and three-dimensional domains 
% increase 
%The strong rise  of the cumulants for all-charge multiplets is increasingly 
%driven by that of  like-sign multiplets, pointing to the influence of Bose-Einstein correlations.
Some of the recently proposed algorithms to simulate Bose-Einstein effects,
 implemented in the Monte Carlo  model \PYTHIA, reproduce reasonably well the measured second- and higher-order
correlations between   particles with the same charge 
as well as those in all-charge particle multiplets.}

\section{Introduction\la{intro}}
Correlations in momentum space 
between hadrons produced in high energy interactions have been
extensively studied over many decades in different contexts.\cite{edw:review}
Being a measure of event-to-event fluctuations of the number of hadrons in a
phase space domain of size $\Delta$, correlations provide detailed information 
on the hadronisation dynamics, complementary to that derived from  
inclusive single-particle distributions and global event-shape characteristics.
The suggestion in\cite{bialas1} that  multiparticle  dynamics might possess (multi-)fractal
properties or be ``intermittent'',  emphasized the importance of studying
correlations as a function of the size of domains in momentum space.
A key ingredient for such studies is the normalized
factorial moment and factorial cumulant technique.
Unlike factorial moments, cumulants of  order $q$ are a direct measure of the stochastic
interdependence among groups  of exactly $q$ particles emitted 
in the same phase space cell.\cite{kendall,Mue71,cumulants} 
Therefore, they are well suited for the study  of true or ``genuine''
correlations between  hadrons and are particularly sensitive to Bose-Einstein correlations.

Two-particle Bose-Einstein correlations (BEC) have been 
observed in a wide range of multihadronic
processes.\cite{review:weiner} 
Such correlations were  extensively studied at
LEP.\cite{opal:bec,delphi:rhoshift,bec:asym}
Evidence for BEC among groups of more than two identical 
particles has also been 
reported.\cite{lep:higherorderbec,otherthanlep:higherorderbec}
The subject has acquired  particular  importance  in connection with 
high-precision  measurements of the $W$-boson  mass 
at LEP-II.\cite{sjo,kh}
For these,  better knowledge of correlations in general is needed,  
as well as realistic Monte Carlo modelling  of BEC.
%\cite{t1,kittel:review}. 

The high statistics OPAL data collected at and near the Z$^0$ centre-of-mass 
energy have been  used to  measure cumulants for   multiplets 
of particles with the same charge,  hereafter referred to as ``like-sign cumulants''.
They are compared to  ``all-charge'' cumulants, 
corresponding to  multiplets comprising particles of any
(positive or negative) charge.
The role of Bose-Einstein-type effects is studied, using recently
proposed BEC algorithms\cite{leif:sjo} implemented in  the  Monte Carlo event generator \PYTHIA\  
for $e^+e^-$ annihilation.\cite{js74}
 Proceeding beyond  the usual analyses of two-particle correlations,
%%%%%
we show that, at least 
within the framework of this model, a good  description can be 
achieved of the factorial cumulants up to fourth order
in one-, two- and three-dimensional phase space domains.

\section{The method\label{fac:method}}
To measure genuine multiparticle correlations in  multi-dimensional
phase space cells, 
we use the technique of normalized factorial cumulant moments, $K_q$, (``cumulants'' for brevity) 
 as proposed in.\cite{cumulants}
The cumulants are computed as in a previous OPAL analysis.\cite{Oic}
A $D$-dimensional
region of phase space  is partitioned into $M^D$ cells of equal size $\Delta$.
From the  number of particles counted in each cell, $n_m$
($m=1,\dots,M^D$), event-averaged unnormalized
factorial moments, $\aver{n_m^{[q]}}$, and  unnormalized cumulants,
$k_q^{(m)}$, are
derived, 
using  the relations given {\eg}
in\cite{kendall}. 
For $q=2,3,4$,  one has
\begin{eqnarray}
k_2^{(m)}&=&\al n_m^{[2]}\ar - \,\al n_m\ar^2,\\
k_3^{(m)}&=&\al n_m^{[3]}\ar - 3\,\al n_m^{[2]}\ar
\al n_m\ar\,
+ 2\, \al n_m\ar ^3\\
k_4^{(m)}&=&\al n_m^{[4]}\ar 
- 4\,\al n_m^{[3]}\ar\, \al n_m\ar
- 3\,\al n_m^{[2]}\ar^2
+12\,\al n_m^{[2]}\ar\,\al n_m\ar^2
-6\, \al n_m\ar^4.
\la{mm}
\end{eqnarray}
Here, $\aver{n^{[q]}}=\aver{n(n-1)\ldots(n-q+1)}$ and 
the brackets $\aver{\cdot}$
indicate that the average over all  events is taken.

Normalized cumulants are calculated using the expression
\begin{equation}
K_q =
({\cal N})^q
\bar{k}_q^{(m)}/
\overline{N_m^{[q]}}.
\la{kmh}
\end{equation}
As proposed  in\cite{kadija:seyboth}, this form 
is used to correct for statistical bias and non-uniformity  of
the single-particle spectra. 
Here, $N_m$ is the number of particles in the $m$th cell summed over all
$\cal N$ events  in the sample,
$N_m= \sum_{j=1}^{\cal N}(n_m)_j$.
The horizontal bar indicates averaging over the $M^D$ cells in each event,
$(1/M^{D})\sum_{m=1}^{M^{D}}$.

Here, data are presented for  ``all-charge'' and for ``like-sign'' multiplets.
For the former, the cell-counts $n_m$  are determined using  
all charged particles in an event, irrespective of their charge. 
For the latter,  the number of  positive particles  and the number of  
negative particles in a cell
are counted  separately. The corresponding cumulants  are then averaged to
obtain those
for like-sign multiplets. 
              
\section{Experimental details\la{data}}
The analysis uses a sample of approximately
$4.1$$\times$$10^6$ hadronic {\z} decays collected from 1991 through 1995.
%About 91\% of this sample was taken  at the {\z}; the
%remaining part has a centre-of-mass energy, $\sqrt{s}$, 
%within $\pm 3$ GeV of the {\z} peak. 
%
The OPAL detector has been described in detail in.\cite{bib-opal}
The results presented  are mainly based on the information from the central 
tracking chambers.
The event selection criteria are based on the multihadronic selection algorithms described in.\cite{Oic}
Multihadron events were selected with at least 5 good tracks,
a momentum imbalance (the  magnitude of the vector sum of the momenta of all charged particles)
 of less than $0.4~\sqrt{s}$ and
the sum of the energies of all tracks (assumed to be pions) greater than 0.2~$\sqrt{s}$.
In addition, the polar angle of the event sphericity axis,
calculated using tracks and clusters 
had to satisfy \mbox{$\mathrm{|\cos \theta_{\mbox{\small sph}}|} < 0.7$} 
in order to accept only events well contained in the detector.  
A total of about $2.3$$\times$$10^6$ events were finally selected for further analysis. 

The cumulant analysis is performed in the kinematic variables
rapidity, $y$,  azimuthal angle, $\phi$, and the transverse momentum variable,
$\ln{p_T}$, all calculated with respect to the sphericity axis.
\begin{itemize}\addtolength{\itemsep}{-1.2ex}
 \item Rapidity is defined  as $y=0.5\ln [(E+p_{\|})/(E-p_{\|})]$, with $E$ and
$p_{\|}$  the energy (assuming the pion mass) and longitudinal
momentum of the particle, respectively. 
Only particles within the central rapidity region $-2.0\leq y\leq 2.0$ were retained.
 \item In transverse momentum  subspace, the  logarithm of $p_T$ is used to eliminate as much as possible the 
strong dependence of the cumulants on cell-size arising from the nearly exponential
shape of the $p_T^2$-distribution. Only particles within the range $-2.4\leq\ln(p_T)\leq0.7$ 
($p_T$ in GeV/{\em c}) were used.
 \item The azimuthal angle  $\phi$ ( $0\leq\phi<2\pi$),
 is calculated with respect to the eigenvector
of the momentum tensor having the smallest eigenvalue in the plane
perpendicular to the sphericity axis. 
 \end{itemize}

\section{Results\label{results}}

\begin{figure}
\begin{center}
\epsfysize=12cm
\epsffile[72 127 472 710]{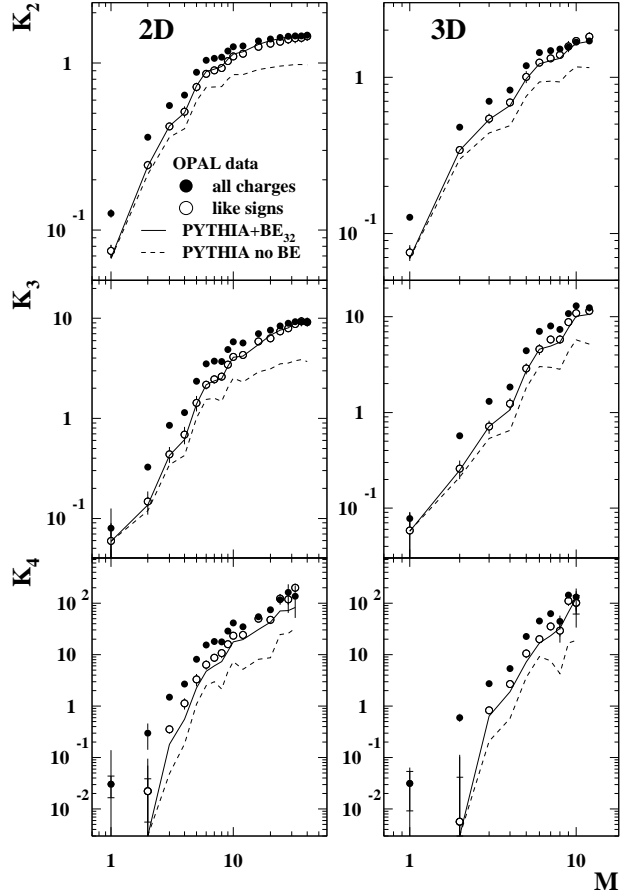} 
\vspace*{-3ex}
\caption{
 The cumulants $K_q$  in 
 two-dimensional $\Delta y\times\Delta\Phi$ (2D) 
and 
three-dimensional $\Delta y\times\Delta\Phi\times\Delta\ln p_T$  (3D) domains
% of rapidity ($y$), azimuthal angle ($\Phi$) and transverse momentum
%($\ \ln \pT$) 
 for all charged hadrons (solid symbols) 
and for  multiplets of like-sign particles 
(open symbols), versus  $M$. 
 Where two error-bars are shown, inner  ones are  statistical, 
and outer ones are
statistical  and 
systematic errors added in quadrature.
The lines  connect   Monte Carlo predictions from \PYTHIA\ without BEC
(dashed) 
and with BEC (full) simulated with algorithm  
$\mbox{BE}_{32}$\protect\cite{leif:sjo} (see text).% 
 } 
\la{fig:2}\end{center}
\end{figure}

\begin{figure}
\begin{center}
%\hs{0.5cm}
\epsfysize=10cm
\epsffile[10 150 530 661]{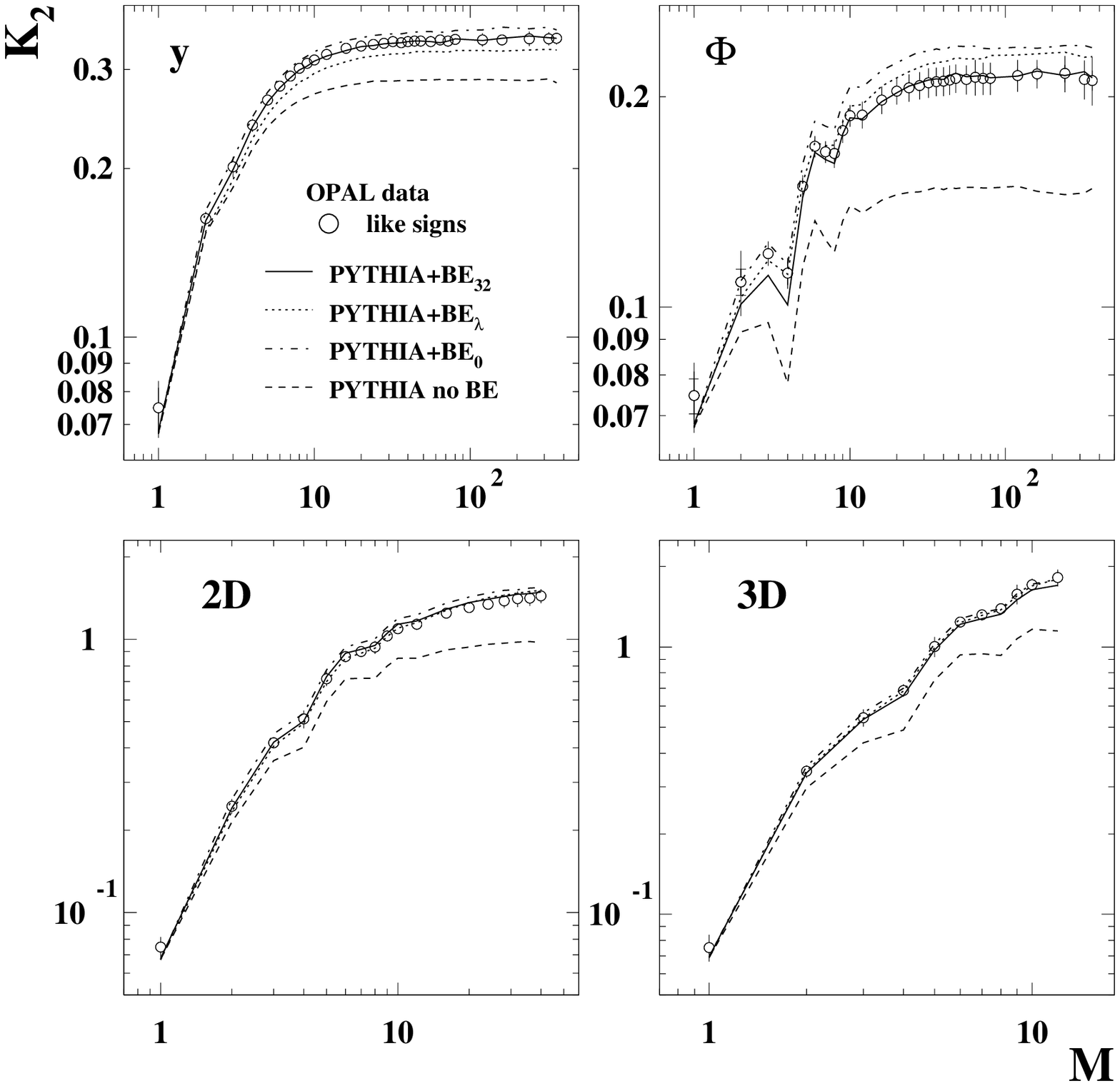} 
\vspace*{-3ex}
\caption{\it  The cumulants $K_2$  for  like-sign pairs
 in one-dimensional domains of rapidity ($y$) and azimuthal angle ($\Phi$), and in
two-dimensional $\Delta y\times\Delta\Phi$ (2D)
 and three-dimensional  $\Delta y\times\Delta\Phi\times\Delta\ln p_T$  (3D) domains 
%of rapidity ($y$), azimuthal angle ($\Phi$) and transverse momentum ($\ \ln \pT$) 
versus $M$.
The error-bars show statistical  and systematic errors added in quadrature.
The lines connect  Monte Carlo predictions from \PYTHIA, without BEC
and with various
Bose-Einstein algorithms\protect\cite{leif:sjo} (see text).  
} 
\la{fig:7}
\end{center}
\end{figure}

%
%\subsection{Like-sign and all-charge cumulants}
The fully corrected normalized cumulants $K_q$ ($q=2,3,4$) 
for all-charge  and like-sign  particle multiplets, 
calculated  in  two-dimensional
$y\times\phi$ (2D) and three-dimensional $y\times\phi\times\ln{p_T}$ (3D) 
phase space cells, are displayed in Fig.~\ref{fig:2}.
It is seen  that
positive  genuine correlations among groups of two, three and four particles are present: $K_q>0$.
Cumulants  in 2D and 3D  continue to increase 
towards small phase space cells. Moreover, the 2D and 3D cumulants are of similar
magnitude  at fixed $M$, 
indicating that  the contribution from correlations in transverse momentum  
is small.
The like-sign cumulants increase faster and
 approach the all-charge ones at large $M$. 
As the  cell-size becomes smaller, the rise of all-charge correlations 
is increasingly driven by that  of  like-sign multiplets.

\begin{figure}
\begin{center}
%\hs{0.5cm}
\epsfysize=12cm
\epsffile[68 128 477 712]{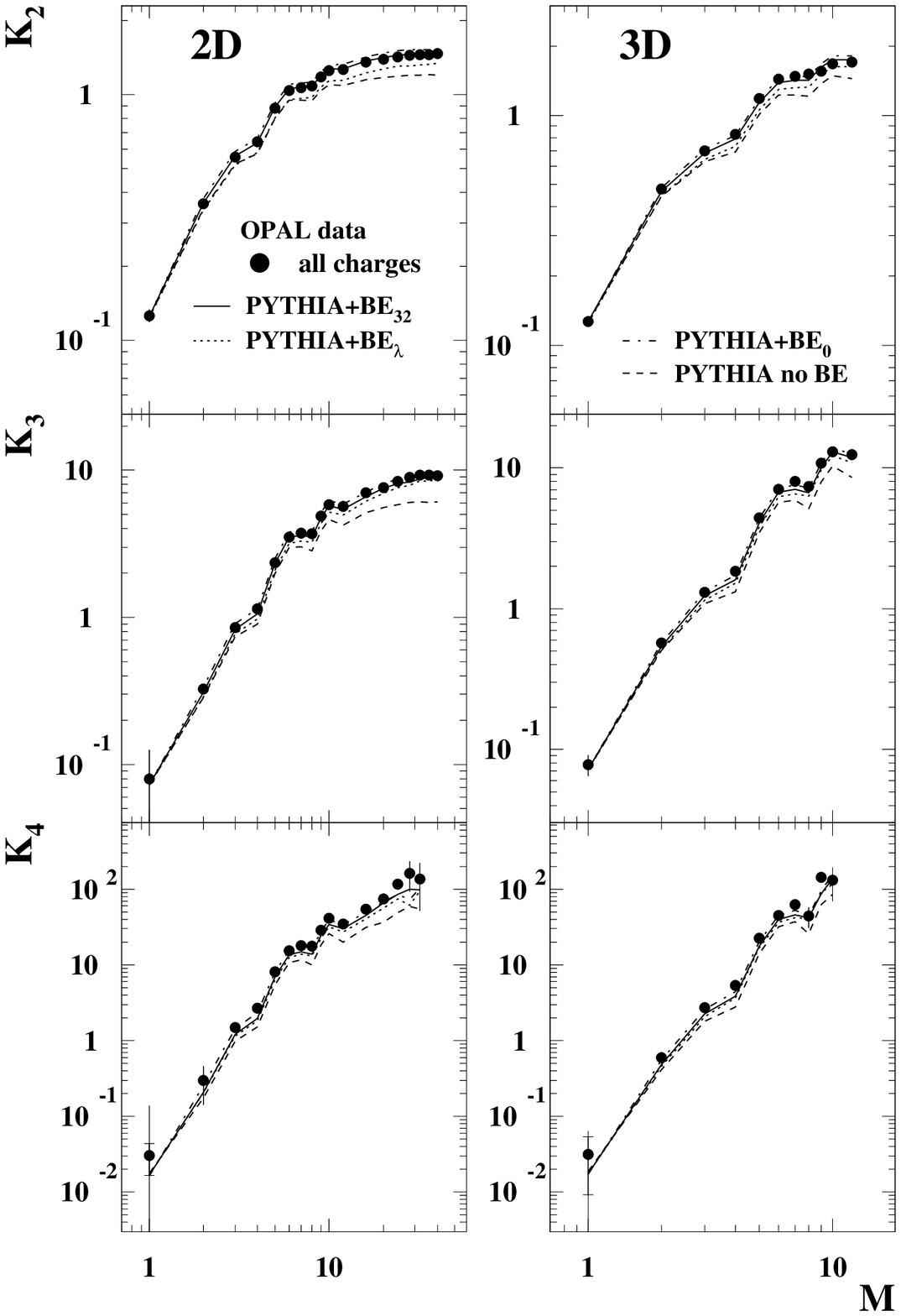} 
\vspace*{-3ex}
\caption{ 
 The cumulants $K_q$   
in  two-dimensional $\Delta y\times\Delta\Phi$ (2D) 
and three-dimensional $\Delta y\times\Delta\Phi\times\Delta\ln p_T$  (3D) domains 
%rapidity ($y$), azimuthal angle ($\Phi$) and transverse momentum
%($\ \ln \pT$) 
for all charged hadrons  versus $M$.
 Where two error-bars are shown, inner  ones are  statistical, 
and outer ones are statistical  and 
systematic errors added in quadrature.
The lines connect  Monte Carlo predictions from \PYTHIA, without BEC
and with various
Bose-Einstein algorithms\protect\cite{leif:sjo} (see text).  
} 
\la{fig:4}
\end{center}
\end{figure}

%\subsection{Model comparison\label{sec:models}}
The cumulant data have been compared with predictions of
the \PYTHIA\  Monte Carlo event generator (version 6.158)  without and with Bose-Einstein effects.
 The model parameters, not related to BEC,  were set at values obtained from a previous tune to OPAL data
on event-shape  and single-particle inclusive   
distributions.\cite{bib-jetset} In this  tuning,  BE-effects were not
included.

To assess the importance  of BE-type short-range correlations between identical 
particles, and their influence on all-charge cumulants, 
we concentrate on the algorithm BE$_{32}$,
described in\cite{leif:sjo}, using parameter values
$\mbox{\tt PARJ(93)}=0.26$~GeV ($R=0.76$~fm) and $\mbox{\tt
PARJ(92)}\equiv\lambda=1.5$.
%
%These values   were determined by varying independently  
%$\mbox{\tt PARJ(93)}$ and $\lambda$ 
%leaving all other 
%model-parameters unchanged, until satisfactory agreement with the 
%measured cumulants $K_2$ for  like-sign pairs  
%was reached.
%
Non-BEC related model-parameters were set at the following values: 
{\tt PARJ(21)}=0.4 GeV,
{\tt PARJ(42)}=0.52 GeV$^{-2}$,
{\tt PARJ(81)}=0.25 GeV,
{\tt PARJ(82)}=1.9 GeV. 
We find that calculations 
with   $\mbox{\tt PARJ(93)}$ in the range $0.2-0.3$ GeV, and the
corresponding
$\lambda$ in the range $1.7-1.3$,  still provide
an acceptable description of the second-order like-sign cumulants.

The dashed  lines in Fig.~\ref{fig:2} are   \PYTHIA\
predictions
for {\em like-sign\/} multiplets for the model without BEC.
Model and  data agree for small $M$ (large phase space domains), indicating
that  the multiplicity distribution in those regions is  well modelled.
However, for larger $M$, the  predicted cumulants are too small.

The solid curves in  Figs.~\ref{fig:2} show 
predictions for {\em like-sign\/} multiplets using the  BE$_{32}$ algorithm.
Inclusion of BEC leads to a very significant improvement of the data description.
Also two-particle and  higher order correlations  in 1D rapidity space are well  accounted for (not shown). 
The predicted 2D and 3D  cumulants  agree well with the data.

The 1D, 2D and 3D cumulants 
for particle pairs with the same charge are  displayed  in Fig.~\ref{fig:7}.
Since BEC occur only
when two identical mesons are close-by in all  three phase space dimensions, 
projection onto lower-dimensional subspaces, such as rapidity 
and azimuthal angle, leads to considerable weakening of the effect. 
Nevertheless, the high precision of the data in    
Fig.~\ref{fig:7} allows to demonstrate  clear sensitivity to the presence 
or absence of BEC in the model.

Whereas the BE-algorithm used implements  pair-wise BEC only, 
it is  noteworthy (see Fig.~\ref{fig:2}) that the procedure 
also induces like-sign higher-order correlations  of approximately correct magnitude.
This seems to indicate that  high-order cumulants are, to a large extent,
determined by the second-order one.

To assess the sensitivity of the 
cumulants to variations in the BEC algorithms available in \PYTHIA,
we have further considered  the algorithms BE$_\lambda$ and BE$_0$.\cite{leif:sjo}
Using the same parameter values as for BE$_{32}$,
we observe that  BE$_\lambda$ slightly  
underestimates $K_2(y)$ and overestimates  $K_2(\phi)$
for like-sign pairs (Fig.~\ref{fig:7}), 
whereas  the results  coincide with those from BE$_{32}$ in 2D and 3D. 
For all-charge multiplets (Fig.~\ref{fig:4}),  
the predicted cumulants  generally fall below those 
for  BE$_{32}$, except for  $K_3$ and $K_4$ in 2D and 3D, where the  differences  are small.
The differences with respect to  BE$_{32}$ are related 
to the different  pair-correlation functions  used in the
algorithms.
Although a different choice of the parameters $R$ and $\lambda$ 
may improve the agreement with the data, we have not attempted such fine-tuning. 

We also considered the predictions based on the  algorithm BE$_0$\cite{leif:sjo}
(dash-dotted curves in the figures) for the same parameter values as quoted above.
For like-sign pairs (Fig.~\ref{fig:7}), $K_2(y)$  and especially $K_2(\phi)$ are overestimated.
%This is also the case for  $K_2(\phi)$  for all-charge pairs  shown in Fig.~\ref{fig:3}.
In contrast, all-charge  higher-order cumulants differ little from those
obtained with   BE$_{32}$.
%
%It should be noted that the  BE$_0$  algorithm, contrary to  BE$_{32}$ and BE$_{\lambda}$,
%enforces   energy conservation  by a global
%rescaling of all final-state hadron momenta. This procedure  affects the full hadronic final state and
%induces a large artificial shift in the $W$-mass when applied to the
%reaction  $e^+e^-\to W^+W^-\to \mbox{hadrons}$. 

To summarize, a comparison with   \PYTHIA\ predictions
shows that 
%besides  correlations due to hard jet production and resonance decays, 
short-range correlations of the BE-type 
are needed, at least in this model,
to reproduce  the magnitude and the $\Delta$-dependence of the cumulants
for like-sign multiplets. This further leads to a much improved description of
the cumulants for  all-charge multiplets.
Since Bose-Einstein
correlations are a well-established phenomenon in multiparticle
production, it is likely that
the above conclusion has wider validity than the model from which it was
derived.

\begin{figure}
\begin{center}
%\hs{0.5cm}
\epsfysize=11cm
\epsffile[15 67 577  756]{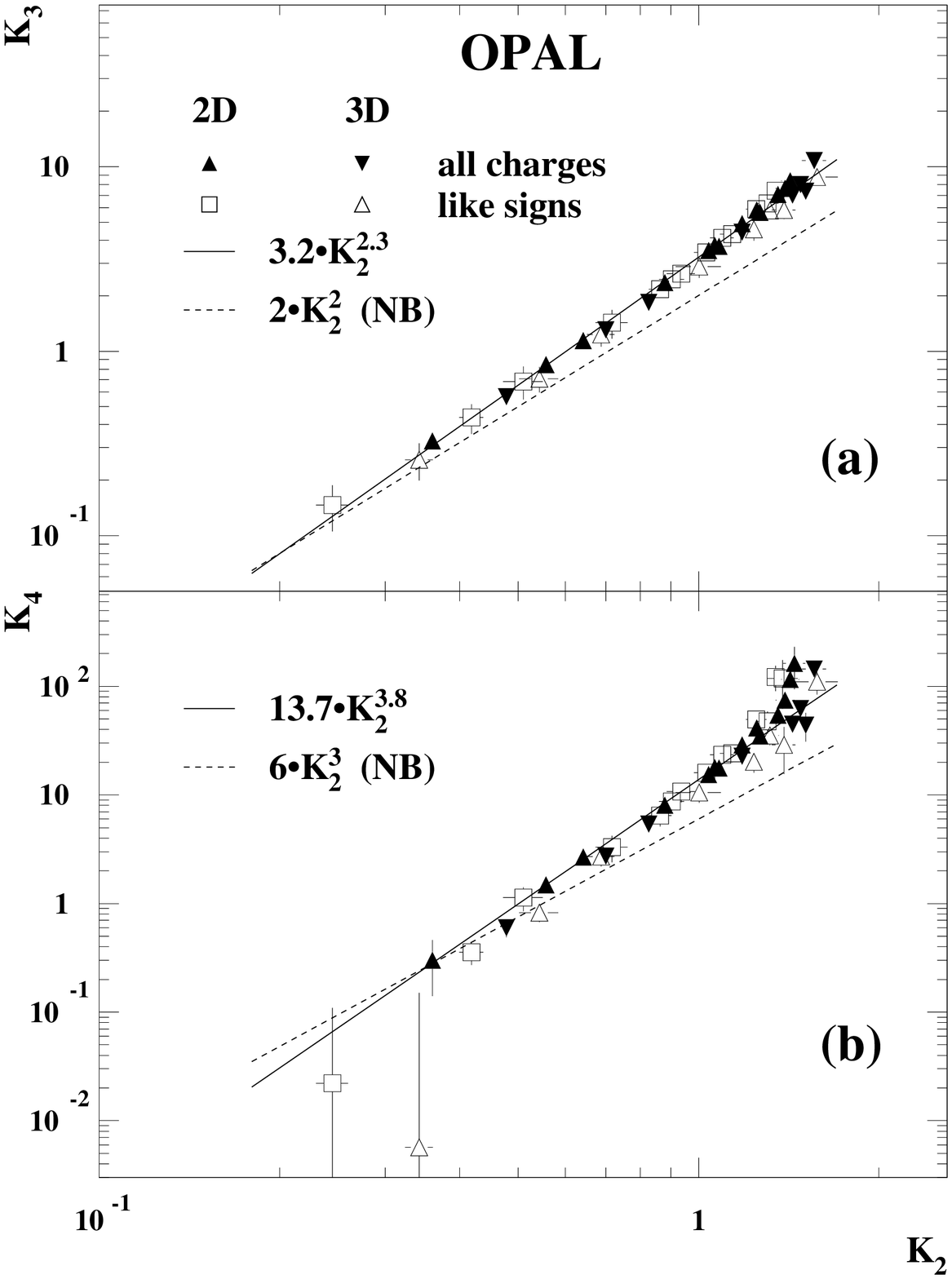} 
\vspace*{-2ex}
\caption{
The Ochs-Wosiek plot 
in two-dimensional $\Delta y\times\Delta\Phi$ (2D) 
and three-dimensional  $\Delta y\times\Delta\Phi\times\Delta \ln p_T$ (3D) 
domains 
%rapidity ($y$), azimuthal angle ($\Phi$) and transverse momentum ($\ \ln \pT$) 
for all charged hadrons (solid symbols) and for multiplets of like-sign
particles  (open symbols). 
The dashed line shows  the function,  $K_q=(q-1)!\,K_2^{q-1}$ ($q=3,4$), 
 valid for
a Negative Binomial multiplicity distribution (NB) in  each phase space cell.
The solid line shows a fit to the relation  $\ln{K_q}=a_q+ r_q\ln{K_2}$.
} 
\la{fig:5}\end{center}
\end{figure}

%\subsection{The Ochs-Wosiek relation for cumulants\label{sec:ochs}}
The success of the \PYTHIA\  model with BEC in predicting both  the magnitude and
domain-size dependence of  cumulants, has led us to consider 
the inter-dependence of these quantities. 
Figure.~\ref{fig:5} shows  $K_3$ and $K_4$   in 2D and 3D, as a function of $K_2$.
The 2D and 3D data for all-charge, 
as well as for like-sign multiplets
follow approximately,  within errors, the same functional dependence. 
The solid lines are a simple fit to the function $\ln{K_q}=a_q+ r_q\ln{K_2}$.
Figure~\ref{fig:5} 
 suggests  that the {\em cumulants\/} of different orders obey  simple so-called ``hierarchical'' 
relations,  analogous to the  Ochs-Wosiek relation,
first established  for {\em factorial moments\/}.\cite{Ochs:Wosiek:relation}
Interestingly,  all-charge as well as   like-sign multiplets are  seen to follow, within errors,
 the  same functional dependence.
%Hierarchical relations of similar type  are commonly encountered, or conjectured, in various branches
%of many-body physics (see {\eg}\cite{cumulants}).

Simple relations among the cumulants of different orders exist for certain
probability distributions, such as the Negative Binomial distribution.\cite{neg:bin}
For this distribution, one has $K_q=(q-1)!\,K_2^{q-1}$ ($q=3,4,\dots$),
showing 
that the cumulants are here solely determined by  $K_2$.
This relation, shown in Fig.~\ref{fig:5} (dashed lines) does not describe the data, indicating  
that the  multiplicity distribution of charged particles, 
and that of like-sign particles, deviates
strongly from a  Negative Binomial in small phase space domains. 

The Ochs-Wosiek type of relation 
exhibited by the data in Fig.~\ref{fig:5} may  explain
why the BE algorithms in \PYTHIA\   generate higher-order correlations of
(approximately) the correct magnitude. 
%Assuming that the hadronization 
% dynamics is such that  higher-order correlation functions can be constructed from
%second-order correlations only,
%methods that are designed to ensure agreement with the two-particle correlation function,
%could
%then automatically generate  higher-order ones of the correct magnitude.

\section{Summary\label{conclusions}}
A comparative study of 
 like-sign and all-charge genuine correlations
between two and more hadrons produced in $e^+e^-$ annihilation at the $Z^0$ energy has been performed
by OPAL using the high-statistics data on hadronic $\mathrm{Z}^0$ decays  
recorded  with the OPAL detector from 1991 through 1995.
%Out of $4.1\times10^6$ events recorded with the OPAL detector, 
%a data  sample of about $2.3\times10^6$ hadronic {\z} decays 
Normalized  factorial cumulants  were measured as a function of the domain size, $\Delta$,  in
$D$-dimensional domains ($D=1,2,3$) in rapidity, azimuthal angle and (the logarithm of) 
transverse momentum, defined  in the event sphericity frame.

Both  all-charge  and like-sign multiplets show
strong positive genuine correlations up to fourth order.
The 2D and 3D cumulants $K_3$ and $K_4$, 
considered as a function of $K_2$, 
 follow  approximately a linear relation of the Ochs-Wosiek  
type: $\ln K_q\sim \ln K_2$, independent of $D$ and the same
for all-charge and for like-sign particle groups.
%This suggests that, for a given  domain $\Delta$,  
%correlation functions of different orders are not
%independent but determined,   to a large extent, by  two-particle correlations. 
%obey a hierarchical structure, meaning that the higher-order correlation functions
%can be expressed in terms of the second-order one.

The \PYTHIA\ model  describes well
dynamical fluctuations in large phase space domains.
However, to achieve a  more satisfactory data description, 
short-range correlations of the Bose-Einstein type 
between  identical particles need to be included.

The Bose-Einstein  model BE$_{32}$ in \PYTHIA\ is  able to simultaneously account for the
magnitude and $\Delta$-dependence of  like-sign 
as well as of all-charge cumulants. The models BE$_0$ and BE$_\lambda$,
when using  the same parameters as for BE$_{32}$, show reasonable agreement with the data.
Although the algorithms implement pair-wise BEC only,
 surprisingly good  agreement with the measured  third-  and fourth-order
cumulants is observed.

%\bibliography{pn-bis,multi}
%\bibliography{pn-bis}

\end{document}